\documentclass[sigconf]{acmart}

\usepackage{enumitem}
\usepackage{soul}
\usepackage{fontawesome5}
\usepackage{wrapfig}

\usepackage{multirow}
\usepackage{soul}
\usepackage{balance}

\newcommand{\cmt}[1]{\ignorespaces}

\AtBeginDocument{%
  \providecommand\BibTeX{{%
    \normalfont B\kern-0.5em{\scshape i\kern-0.25em b}\kern-0.8em\TeX}}}

\copyrightyear{2022}
\acmYear{2022}
\setcopyright{acmlicensed}\acmConference[CHIWORK]{2022 Symposium on Human-Computer Interaction for Work}{June 8--9, 2022}{Durham, NH, USA}
\acmBooktitle{2022 Symposium on Human-Computer Interaction for Work (CHIWORK), June 8--9, 2022, Durham, NH, USA}
\acmPrice{15.00}
\acmDOI{10.1145/3533406.3533418}
\acmISBN{978-1-4503-9655-4/22/06}

\begin{document}

%%
%% The ``title'' command has an optional parameter,
%% allowing the author to define a ``short title'' to be used in page headers.
\title[A Bottom-Up End-User Intelligent Assistant Approach to Empower Gig Workers]{A Bottom-Up End-User Intelligent Assistant Approach to Empower Gig Workers against AI Inequality}

% dakuo suggested title:
% \title{``Everyone Can Build a Story Buddy'': Towards a Human-in-the-Loop Framework to Automatically Generate Conversational Agents to Support Parent-Child Interactive Storytelling}

% \title{``Every Parent Can Build a StoryBuddy'': Towards a Human-in-the-Loop AI System to Enable Parents Build Conversational Agents to Support Parent-Child Interactive Storytelling}
%%
%% The ``author'' command and its associated commands are used to define
%% ``authornote'' and ``authornotemark'' commands
\author{Toby Jia-Jun Li\textsuperscript{1}, Yuwen Lu\textsuperscript{1}, Jaylexia Clark\textsuperscript{1}, Meng Chen\textsuperscript{1}, Victor Cox\textsuperscript{1}, Meng Jiang\textsuperscript{1}, Yang Yang\textsuperscript{2}, Tamara Kay\textsuperscript{1}, Danielle Wood\textsuperscript{1}, Jay Brockman\textsuperscript{1}}
 \email{{toby.j.li, ylu23, jclark25, mchen24, vcox2, mjiang2, tkay, dwood5, jbb}@nd.edu, yyang87@syr.edu}
\affiliation{%
  \institution{\textsuperscript{1}University of Notre Dame, \textsuperscript{2}Syracuse University}
   \city{}
   \state{}
  \country{}
}

\renewcommand{\shortauthors}{Li et al.}

\begin{abstract}

The growing inequality in gig work between workers and platforms has become a critical social issue as gig work plays an increasingly prominent role in the future of work. The AI inequality is caused by (1) the \textit{technology divide} in who has access to AI technologies in gig work; and (2) the \textit{data divide} in who owns the data in gig work leads to unfair working conditions, growing pay gap, neglect of workers’ diverse preferences, and workers’ lack of trust in the platforms. In this position paper, we argue that a bottom-up approach that empowers individual workers to access AI-enabled work planning support and share data among a group of workers through a network of end-user-programmable intelligent assistants is a \textit{practical} way to bridge AI inequality in gig work under the current paradigm of privately owned platforms. This position paper articulates a set of research challenges, potential approaches, and community engagement opportunities, seeking to start a dialogue on this important research topic in the interdisciplinary CHIWORK community.\looseness=-1

\end{abstract}

\begin{CCSXML}
<ccs2012>
<concept>
<concept_id>10003120.10003121</concept_id>
<concept_desc>Human-centered computing~Human computer interaction (HCI)</concept_desc>
<concept_significance>500</concept_significance>
</concept>
<concept>
<concept_id>10002951.10003227.10003241</concept_id>
<concept_desc>Information systems~Decision support systems</concept_desc>
<concept_significance>300</concept_significance>
</concept>
<concept>
<concept_id>10003120.10003130.10003131.10003570</concept_id>
<concept_desc>Human-centered computing~Computer supported cooperative work</concept_desc>
<concept_significance>300</concept_significance>
</concept>
</ccs2012>
\end{CCSXML}

\ccsdesc[500]{Human-centered computing~Human computer interaction (HCI)}
\ccsdesc[300]{Information systems~Decision support systems}
\ccsdesc[300]{Human-centered computing~Computer supported cooperative work}

\keywords{gig workers, AI inequality, intelligent assistants, human-AI collaboration}


\begin{teaserfigure}
        \centering
        \includegraphics[width=0.75\linewidth]{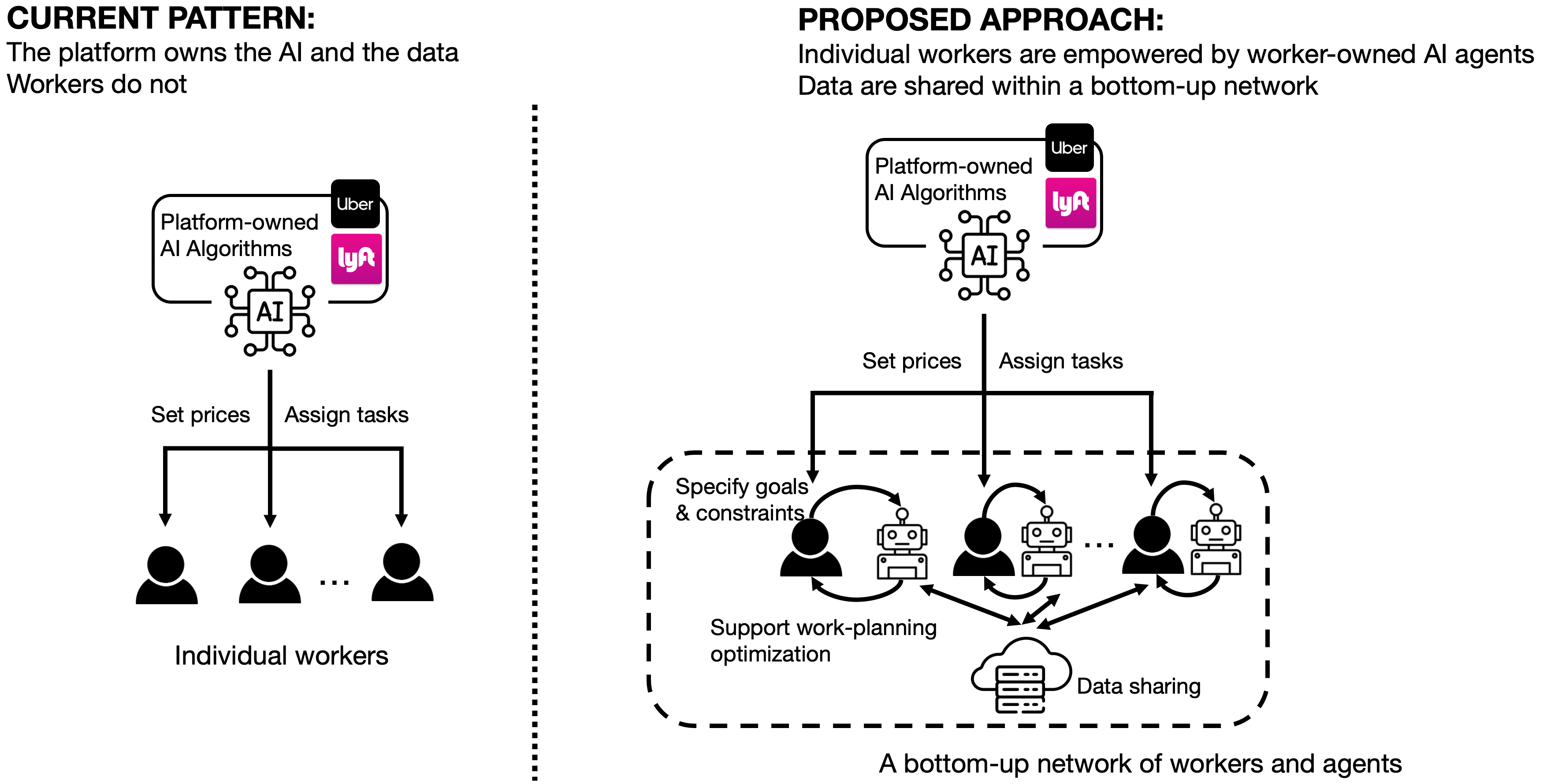}
        \caption{Comparing the current platform-worker pattern with the proposed approach}
        \label{fig:approach}
    \end{teaserfigure}

\maketitle

\section{Introduction}
Digitally-mediated gig work, where individual workers provide on-demand work (e.g., rideshare) through online platforms (usually on mobile apps), constitutes a significant and growing portion of the workforce. Reports estimated that about one-third of the U.S. workforce participate in the gig economy~\cite{mitic_24_2021}, while around one percent of the U.S. workforce (1.6 million people) directly work on app-based gig work platforms~\cite{usbls_2017, jackson2017rise, farrell2018online}. The importance of gig work is expected to continue to rise in the future of work~\cite{gray2019ghost}. The labor participation in the gig economy skyrocketed during the COVID-19 pandemic---prior research found a 25\% increase in average labor supply on gig economy platforms in response to the economic impact of COVID-19~\cite{cao2020impact}. The efforts to empower gig workers have drawn a lot of attention due to recent high-profile events such as the pass of CA Prop. 22 that regulated the roles and benefits of app-based gig workers~\cite{california_attorney_general_proposition_2021}, the repeal of rules by the U.S. Dept. of Labor to prevent ``employer misclassification'' of workers~\cite{segal_labor_2021}, and the strikes and protests of gig workers around the nation~\cite{gig_worker_rising_gig_2021}. 
% \tlcomment{introduce the ai inequality}

Compared to conventional contingent work, a key unique characteristic of digitally-mediated gig work is the ubiquity of artificial intelligence (AI) involvement. For example, in rideshare, AI systems match drivers with customers and determine drivers' pay. However, \textbf{the use of AI contributes to the widening inequality in gig work} between platforms and workers: these AI systems are ``owned'' by platforms and designed to \textbf{optimize for platforms' best interests}. In contrast, individual gig workers' unique goals, preferences, and constraints are often neglected in the existing gig work platforms. The AI systems in these platforms lack transparency and interpretability for users to understand the algorithmic recommendations~\cite{cameron_this:_2021, dillahunt_sharing:_2017, gray2019ghost}. Many platforms are also known to have systematic gender, racial, and socioeconomic biases in their AI algorithms~\cite{hannak_bias:_2017, thebault-spieker_toward:_2017, thebault-spieker_avoiding:_2015, dillahunt_reflections:_2017}. Meanwhile, high technical barriers prevent workers from accessing AI technologies that work in their best interests~\cite{lieberman2006end}, leading to a \textbf{technology divide}~\cite{warschauer2004technology}. It requires significant expertise to ``program'' an AI system that can model complex gig work situations and provide work planning support that accommodates the diverse needs of workers. \looseness=-1

% \tlcomment{talk about the data imbalance issue}

%% Why not design a more equitable algorithm? 

Another key contributing factor to the AI inequality in gig work is the \textbf{data divide} in who has access to and control over data~\cite{delacroix2019bottom,andrejevic2014big}. Simply put, gig work platforms have access to data from all drivers as well as the technologies to use those data while drivers do not.  Platforms, through their applications, collect and access data from all drivers and customers, which are used to model worker behaviors, task characteristics, and customer demands. Workers do not have access to such collective data---each individual worker, at most, can only self-track the data of themselves. Therefore, even if workers had access to the technology, their \textbf{lack of data access} would prevent them from using AI systems for work planning assistance in their best interest. This issue has recently been recognized as a key challenge toward a more equitable future of gig work through initiatives such as Digital Worker Inquiry~\cite{edinburgh_centre_for_data_culture__society_digital_2022}.    
    
In this position paper, we argue that a practical approach to bridge the AI inequality in gig work is through the design, development, and deployment of a bottom-up network of end-user intelligent assistants. Each assistant is paired with a worker, collects work-related data from them, and shares the data within the network to close the data divide. With the available data, the assistant can model and predict the market and empower workers to optimize their work according to their personal goals, preferences, and contexts, reducing the technology/AI divide. 

The envisioned approach is translational with the potential to make immediate impacts on gig worker communities within the current paradigm of privately-owned gig work platforms. Compared with other directions e.g., developing more equitable pricing and task allocation algorithms for the platform, developing a more explainable and transparent worker's app for the platform, the technology required for this approach is worker-centric without requiring any cooperation, access to additional data, or software access from the platform. This is important to the practicality of the approach, as we do not expect platforms to be completely cooperative in the effort to reduce AI inequality in gig work because it might work against their financial interests. 

This research agenda can also, through the deployment and data collection process, contribute to fundamental understandings in characterizing and measuring AI inequality in gig work and explore worker needs and strategies in effective human-AI collaboration between gig workers and AI-enabled assistants. For broader impacts, these findings can raise awareness of AI inequality in gig work, providing empirical evidence and implications for labor advocacy, worker movements, and legislative and policy efforts in market regulation and worker rights protection in gig work. Beyond making immediate direct impacts within the current gig work paradigm of privately-owned platforms, findings from this approach (e.g., how workers collaborate with each other through data sharing and AI-enabled facilitation in a decentralized network) may also contribute to the foundation for a long-term direction in designing new community-owned decentralized gig work platforms, an idea that many in the research community are excited about.

%\tlcomment{TODO: summarize our approach}
    
\section{Background}
\subsection{The Role of Gig Work Platforms}
An important factor to consider for AI inequality specifically in gig work is the role of gig work platforms~\cite{vallas2020platforms}. From a labor economics point of view, the main purpose of these platforms is to reduce ``search frictions'' by helping match those looking for workers (e.g., riders in ridesharing) and those looking for work (e.g., rideshare drivers)~\cite{pissarides2000equilibrium, prassl2018humans,hoffmann2019fairness}. The platform plays an \textit{intermediary} role where it matches workers with customers and sets prices using data and algorithms they have~\cite{prassl2018humans,hoffmann2019fairness,wood2019good}. 
    
While the platform provides values as an intermediary (and provides other services e.g. mechanisms for quality control and infrastructures for handling payments), it takes unfair advantages from the lack of transparency in the matching and the pricing process through its \textit{monopoly} on data and algorithms~\cite{van2020platform,hoffmann2019fairness,rosenblat2016algorithmic}: the platform has exclusive access to the data of labor supplies and demands and the predictive model of prices that each party is willing to accept. When a match (e.g., between a rider and a driver) is made, neither party knows the price that the other party is paying/receiving, nor information of the other possible riders and drivers in the market. This allows the platform to optimize its task assignment and pricing algorithms in the best interests of the platform~\cite{rosenblat2018uberland,wood2019good,rosenblat2016algorithmic}. The proposed bottom-up end-user intelligent assistant approach seeks to bridge the AI inequality in gig work by breaking the platform's monopoly on data and algorithms, enabling workers to utilize data-driven AI assistance that works in their best interests.\looseness=-1 

\subsection{Prior Studies of Inequality in Gig Work}
Previous work has demonstrated that gig work is different from work in the formal economy in three major ways. First, gig work is precarious with individuals unable to control how many ``gigs'' will be available and whether they will be able to accept each incoming gig. Second, gig work is typically low-wage labor~\cite{prassl2018humans}. Many sociologists argue that gig work depresses wages in order to create exploitative conditions that make workers increasingly reliant on gig work~\cite{graham2018towards,prassl2018humans}. Lastly, gig workers are not considered employees and thus are not protected by federal employment law~\cite{graham2020fairwork,graham2018towards,prassl2018humans}. Overall, previous research shows that gig work depresses wages, is precarious, and is unprotected. \looseness=-1

Inequalities in gig work are ubiquitous. For example, a 2017 study reported a significant correlation between perceived gender and race of workers with worker evaluations, which harms employment opportunities of workers~\cite{hannak_bias:_2017}. Aside from worker evaluations, which are more subjective, systematic biases are still prevalent---Most physical gig work tasks are geographic in nature. Geographic principles interact with platform designs to create systematic biases in which services like Uber are more effective in dense, high socioeconomic status (SES) areas than in low-SES areas~\cite{thebault-spieker_toward:_2017}, similar to patterns previously found in peer-production contents like Wikipedia and OpenStreetMap~\cite{johnson_not_2016}. A 2019 study identified unfair income distributions on a ride-hailing platform between ``successful`` and ``unsuccessful drivers''~\cite{suhr_two-sided:_2019}. Previous studies reported strong tension between the two parties due to workers' struggle against power asymmetries maintained by the platform, since they are marginalized in the decision-making process of the platform design~\cite{ma_stakeholder_2018, rosenblat2016algorithmic,rosenblat2016algorithmic}. The lack of transparency and control resulted in a feeling of dehumanization for workers~\cite{mohlmann2017hands}. These tensions have led to worker movements in gig work ecosystems~\cite{kinder_gig_2019}. 

 % We are also going through a fundamental shift in the worker-AI relationship---In the past years, AI systems have been playing \textbf{a managerial and supervisory role}, leaving little agency for gig workers. However, with recent worker movements and legislature changes (e.g., CA Assembly Bill 5~\cite{california_legislative_information_bill_2019}), algorithmic systems are shifting to \textbf{a partner role}. Uber drivers are now able see more information (e.g., the destination) of a ride before accepting it, for instance. With the new agency, workers need algorithmic systems that work in their best interests to help them with work-related planning and decisions. \looseness=-1

While toolkits and techniques have been developed to help AI system owners and operators to make their algorithms fairer in a top-down fashion~\cite{bellamy2019ai,holstein_2019_improving}, it is unclear how much gig work platforms have adopted them due to the power inequality between the platform and workers. Therefore, our proposed approach seeks to address the inequality and biases in gig work using a \textit{bottom-up} approach instead, empowering each individual worker with better AI-enabled decision-making and planning assistance that works in their best interests towards their unique personal goals. 

\subsection{Intelligent Assistants for Work Planning and Optimization}
The end-user intelligent assistant we envision in this position paper will be an instance of \textit{end-user programmable} systems that allow users without technical expertise to ``program'' them to help users with decision-making, planning, and automation. Research on intelligent assistants that \textit{collaborate} with users in the planning and decision-making of complex problems has been around for decades since early systems like \textsc{Trips}~\cite{ferguson1998trips} and \textsc{Trains}~\cite{ferguson1996trains}. A key approach for effective human-AI interaction in such assistants is to support \textit{mixed-initiative interaction}~\cite{horivitz1999principles,allen1999mixed}, where an agent infers the goals and needs of users with uncertainties, employs multi-turns of interactions with users to resolve key uncertainties, and provides information that \textit{augments} the user's capability to make decisions instead of making decisions for the users. Several existing intelligent assistants such as~\cite{li_pumice:_2019,li_sugilite:_2017,li_interactive:_2020,li_sovite:_2020} have been designed and developed in the domain of general-purpose task automation on smartphone apps without a particular focused work domain. \cmt{We have also conducted prior work in this space but in the domain of general-purpose smartphone app task automation~\cite{li_pumice:_2019,li_sugilite:_2017,li_interactive:_2020}.} Two long-standing challenges in mixed-initiative assistants are particularly relevant to the approach we discuss in this position paper: (1) enabling workers without technical expertise to specify their goals and needs; and (2) effectively explaining the assistant's results and recommendations in the context of use. 
    
Prior studies~\cite{trippas_2019_learning} identified the needs for work-related task planning support by office information workers. Various intelligent assistants have also been built to support the planning and optimization of specific work tasks such as meeting scheduling~\cite{cranshaw_calendar_2017} and writing~\cite{greer2016introduction}. Recent initiatives such as Driver's Seat~\cite{drivers_seat_cooperative_drivers_2022} piloted a co-op approach where drivers voluntarily contribute their data and an app help drivers analyze the data to answer questions such as the most productive time to work, the best platform to work on, and whether the driving strategy is working. However, Driver's Seat uses a centralized approach (that relies on selling worker-contributed data to third parties) and provides limited AI-enabled planning and optimization support to drivers. To achieve the vision in this position paper, as a research community, we need to investigate the unique goals, needs, preferences, and work contexts of gig workers and design new features, interfaces, and interaction strategies for a personal intelligent assistant to address them.

\section{Formative Study Summary}
We have conducted several prior studies on inequality in gig work that informed the objectives of this research. Through a collaboration with New York Worker Institute, we analyzed online reviews from digitally-mediated platform gig workers addressing their work conditions. We found that the negative impact of automated matching between workers and consumers was a major source of complaints from gig workers. In another formative study, we quantitatively analyzed the relationship between racial residential segregation, income inequality, and labor participation in the platform economy across all U.S. counties between 2010 and 2018 using a fixed effect panel model~\cite{halaby2004panel}. The result, on a national level, confirms the findings of previous studies on the negative impact of the designs of the gig work platform algorithms on the inequality between high-SES and low-SES areas and workers from different gender and racial groups~\cite{hannak_bias:_2017,suhr_two-sided:_2019,thebault-spieker_avoiding:_2015}.  The results of both studies provide evidence for the negative impact of AI inequality on workers in existing gig work platforms.\looseness=-1

\paragraph{Worker perception of AI inequality} To inform the proposed approach, we have conducted six semi-structured interviews with rideshare drivers and two brainstorm sessions with labor organizers at Chicago Rideshare Advocates~\cite{chicago_gig_alliance,chicago_rideshare_advocates}. The findings of these studies confirmed workers' interest in using an intelligent assistant owned by workers to help them plan their work and joining a bottom-up data-sharing network of drivers to bridge the ``data divide'' in gig work. Workers reported anecdotes of how they \textit{suspected} the platform-owned task allocation and pricing algorithms were working ``against them'' to maximize the profit of the platform by e.g., preventing them to hit bonuses by stopping assigning them tasks when they are close to a milestone (Platforms like Uber sometimes run promotions that award drivers a bonus when they complete \textit{X} tasks that fulfill certain criteria within a given time frame) and charging high surge prices from customers without raising workers' pay. Some workers have also reported the use of some primitive methods to share information between themselves, such as messaging each other on platforms such as WhatsApp or Facebook Messenger to share work-related information e.g., a high demand at the airport or anecdotes of high surge pricing near some areas.\looseness=-1

\paragraph{Unmet worker needs} Through these engagements, we have identified challenges in lowering the barrier of use, making the algorithmic results interpretable, and accommodating the diverse needs and contexts of workers. Although most gig workers, including rideshare drivers, are legally independent contractors, the algorithmic management systems in existing gig work platforms provide workers with little space in \textit{choosing} tasks that they want to do according to their preferences, constraints, and personal values. For example, workers are not able to view \textit{all} the available tasks on the market. Instead, the platform pushes individual task recommendations to them for which workers need to either accept or decline on the spot. In most markets (some states or cities mandate worker access to task information when making decisions), workers cannot view important characteristics of the task (e.g., ride destination, pay amount) either prior to accepting and will even be ``punished'' by the algorithm for decline too many tasks sometimes. This ``recommendation'' or ``task allocation'' algorithm is completely opaque to the workers---workers have no means to configure it according to their unique needs or goals. 

The workers reported a wide range of goals, constraints, and work contexts currently not accommodated by the platform. For example, a worker had significant caregiving duties at home, therefore she preferred to stay within a certain distance when driving for the rideshare service so she could get home quickly if needed. Goal-wise, most workers probably see to maximize their monetary profit i.e., the expected pay from the task minus the expected cost (e.g., gas for rideshare drivers), but some drivers might have different personal values e.g., optimizing the carbon emission or devaluing tasks that involve being stuck in traffic, even if they may pay well, for emotional reasons. Lastly, some work contexts also contribute to personal value, but are neglected by existing platform algorithms. For example, there are drivers who drive hybrid cars that have better gas mileage on local streets than on highways, which is the opposite of most other vehicles. but this factor was likely completely neglected in the platform's algorithm for task allocation.\looseness=-1

\section{The Envisioned Approach}
As shown in Figure~\ref{fig:approach}, the key technical component in the approach that we envision is a bottom-up network of intelligent assistants. Each gig worker runs a Gig Worker Assistant app locally on the same phone used to run the gig work platform. The worker specifies their unique personal values, constraints, and work contexts in the assistant. The assistant can collect task information (e.g., the time needed, distance, proposed pay) of each task recommended by the gig work platform from the GUI of the gig work platform app. The worker also (optionally) shares their real-time location with the server. In our vision, this network of intelligent assistants should achieve the following goals:

\begin{enumerate}
        \item Collect the worker's work-related data and share it with the assistants of other workers in the same network in real-time;
        \item Empower the gig worker to specify their unique goals, constraints, and context;
        \item Model worker behaviors and customer demands using data shared in the network and help the worker optimize work planning based on their own goals, constraints, and context;\looseness=-1
        \item Provide the worker with decision-making support in an explainable interface that fits into the worker's existing work context.\looseness=-1
\end{enumerate}

\subsection{The Bottom-Up Approach for Bridging the Data Divide}
The adoption of AI in existing gig work platforms is \textit{top-down}, leading to the data divide between platforms and workers. Currently, a centralized AI model, owned by the platform, collects data from all workers and customers and then uses the data to make algorithmic decisions, such as pricing and task assignments. The lack of access to data for workers prevents them from having AI that ``works for them'', contributing to the AI inequality. 
    
In response, our envisioned approach is bottom-up and worker-centric without requiring any additional data or software access from the platform. As shown in Figure~\ref{fig:approach}, each worker has an intelligent assistant. A network of assistants from a group of workers share data that each assistant collects of task supply, task demand, and task characteristics within the network, allowing predictive modeling of the market that enables work-planning support for gig workers. This bottom-up network approach does \textit{not} require the buy-in of the platform, scales up easily, and allows flexible configurations and preferences from workers. 

\begin{figure*}[tb]
        \centering
        \includegraphics[width=4.8in]{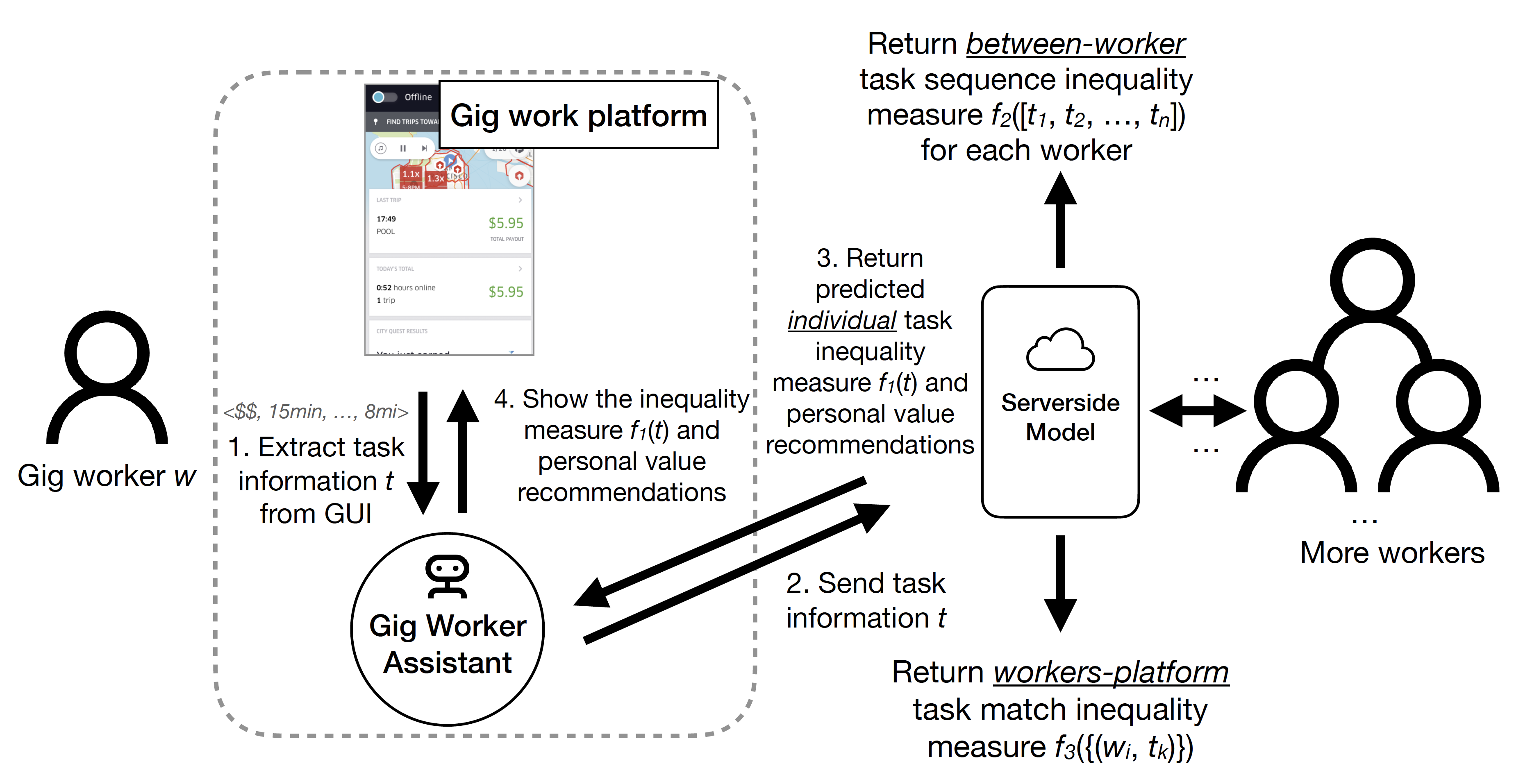}
        \caption{The workflow of the envisioned Gig Work Assistant}
        \label{fig:architecture}
\end{figure*}

% \tlcomment{discuss the privacy issue too?}

\subsubsection{Predictive Modeling User Behaviors and Task Characteristics}
% \tlcomment{TODO: discuss the predictive model}

Computational models empowered by AI and machine learning can capture complex patterns in human behavior and make accurate predictions in many applications~\cite{jiang2014fema,jiang2015general,jiang2016little}.
The objective of predictive modeling in the envisioned approach is to create computational models that use the collected (both current and past) data of a smaller group of workers' behaviors and task characteristics, such as job duration, source/destination locations, price, and task acceptance/declining to predict those of the entire market in real-time. Predictive models will be deployed to provide recommendations to workers for planning optimization.\looseness=-1

Previous work on representation learning models for user behavioral modeling (e.g., \textsc{CalendarGNN}~\cite{wang2020calendar}) used graph neural networks for automatically learning vector representations (or known as ``embeddings") of users from their spatiotemporal behavior data. The architecture of such models incorporates two networked structures. One is a tripartite network of items, sessions, and locations. The other is a hierarchical calendar network of hour, week, and weekday nodes. It first aggregates embeddings of location and items into session embeddings via the tripartite network and then generates user embeddings from the session embeddings via the calendar structure. The user embeddings preserve spatial patterns and temporal patterns of a variety of periodicities (e.g., hourly, weekly, and weekday patterns).

\textsc{CalendarGNN} was shown to be effective in identifying user preferences and predicting their behavior on social media sites, online shopping websites, and news reading mobile applications. A potential challenge of applying \textsc{CalendarGNN} and many other machine learning models on task demand prediction for gig workers is the sparsity, noise, and uncertainty of the collected data. Interpretable and robust machine learning models are needed for reliable user behavior modeling in this domain.\looseness=-1

\begin{figure}[b]
            \centering
            \includegraphics[width=1.9in]{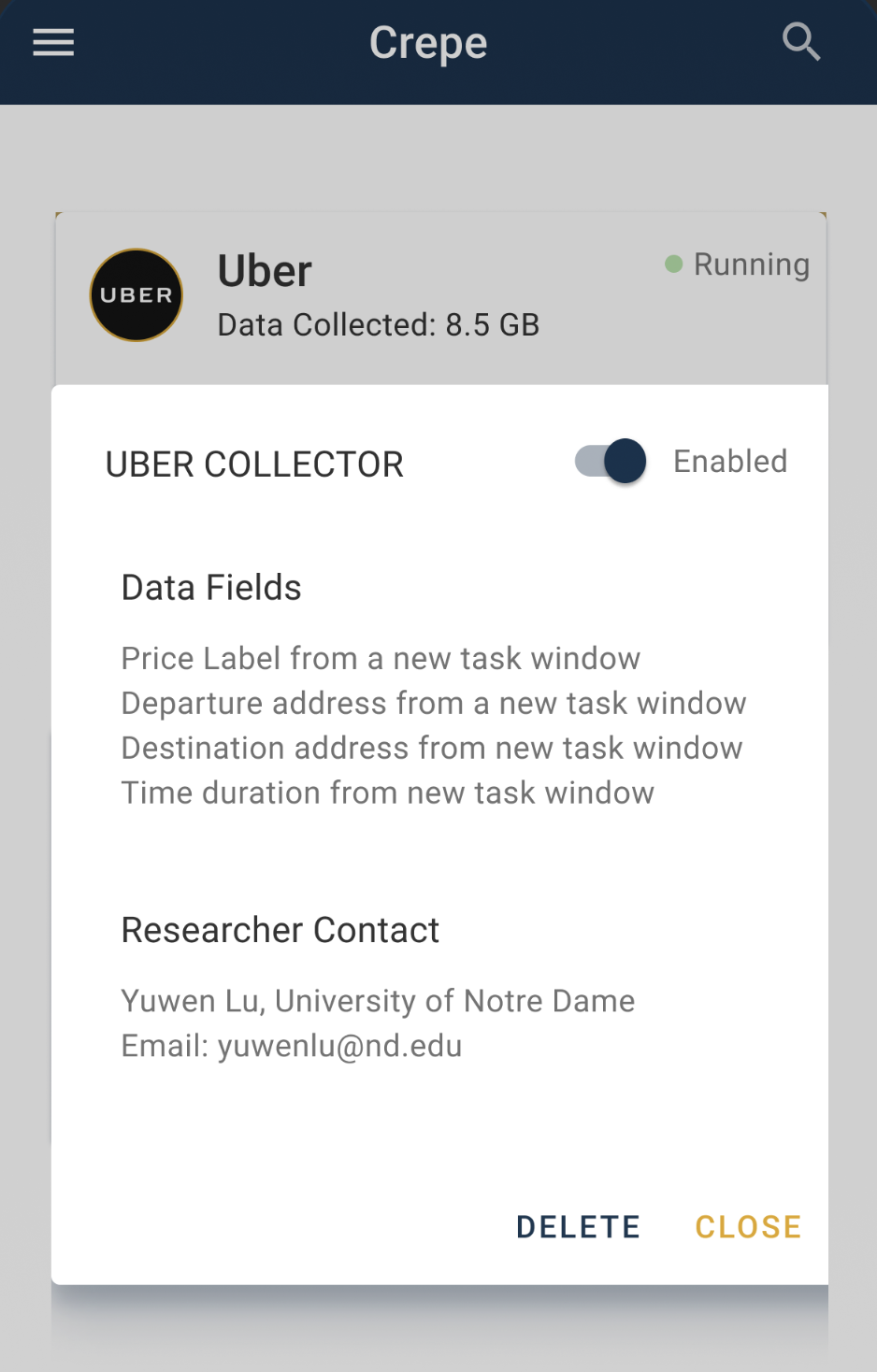}
            \caption{A screenshot of the data crawler \textsc{Crepe}}
            \label{fig:collector_screen}
\end{figure}

\subsubsection{Techniques for Data Crawling}
To show the feasibility of this approach, we have already developed and tested
a practical technique for crawling task information from the GUIs of existing gig work platforms. A crawler (named \textsc{Crepe}) runs on each worker's phone as an Android background service that uses the GUI-based instrumentation method, which extracts data directly from the GUIs of existing apps of gig work platforms. Whenever a piece of information is displayed in the gig work platform for the worker to see, the crawler can extract it. It also listens to the worker's input into the platform app (e.g., accepting a task). This method provides compatibility with different gig work platforms and different makes and models of devices (Android only) that workers use.

This GUI-based approach for data crawling in third-party apps uses a graph-based model presented~\cite{li_appinite:_2018} that grounds a semantic entity (e.g., the estimated pay of a gig task) to a relational query on a graph representation of a third-party app GUI, allowing robust retrieval of the entity value from the GUI in future contexts. It can also embed the semantics of higher-level task intents (e.g., compare the pay of two gig work sessions) of GUI-based tasks.

A limitation of the GUI-crawling approach is that it cannot collect the customer payment needed for modeling \textit{worker-vs-platform biases}, as the customer-paid price is not visible to workers in most gig work platforms. However, this limitation can be complemented by a simulated query method~\cite{thebault-spieker_avoiding:_2015}. When a worker receives a task, a back-end server can simulate a GPS location at the starting location of the task, run the corresponding customer's app (e.g., the customer version of the Uber app), and obtain a price quote for an identical task. This method allows the assistant's model to estimate the user payment for a task using the actual pricing algorithm (which is often kept a secret by the platform). 

% \tlcomment{techniques for data crawling}

\subsection{Empowering Workers with End-User-Programmable Assistants}
As summarized earlier, in our envisioned approach, an end-user-programmable assistant will be paired up with each gig worker and empower them with AI-enabled task planning and decision-making support. This assistant is
end-user-focused in three ways: it is \textbf{owned} by an individual worker, is \textbf{programmed} by an individual worker, and \textbf{communicates} its results \textbf{directly} to an individual worker. The worker ``teaches'' the assistant their unique goals, constraints, and contexts (which form the \textit{personal value} of workers). The assistant then utilizes the predictive model of the market, made with data shared by the assistants of other drivers within the network, to provide workers with suggestions on whether to accept a task, the areas to go to for the next task, and the hours to work that optimize for the worker's personal value. The ownership gives the worker complete control over the assistant and ensures that the assistant works in the worker’s best interest, unlike the task recommendations provided by platform-owned algorithms. Designing such an assistant presents the following research opportunities and challenges:

\paragraph{Specifying worker goals, constraints, and contexts} For the assistant to determine the \textit{personal value} of a recommended task for a worker, the worker needs to configure the assistant by specifying their unique goals, constraints, and context. Because most of the workers do not have significant technical expertise, it is likely not feasible for them to directly define and configure these parameters for the assistant. A system needs to guide them through the process. The main challenge is twofold: the system has to first help the worker resolve their own \textit{uncertainty} and figure out what they want through a \textit{mixed-initiative}~\cite{horivitz1999principles} process. Then, the system should enable the worker to express their intents in a formalized way that the assistant can understand and execute. We expect that a combination of multiple interaction modalities and strategies~\cite{sarmah_geno:_2020,oviatt_ten_1999}, such as natural language programming~\cite{lieberman2006feasibility,lieberman2006end}, direct manipulation~\cite{hutchins1985direct}, and programming by example~\cite{lieberman_end-user_2006,cypher_watch_1993} techniques would be useful here.\looseness=-1

\paragraph{Displaying and explaining results} The assistant should display the result of work planning assistance in a way that is appropriate for the background and the work context of gig workers. Any recommendations for accepting or declining a task or going to a new area etc. should be accompanied by explanations of why it contributes to the worker's personal value in an easy-to-understand format. The format should also accommodate the worker's unique context of use. For example, rideshare drivers may need to be able to consume such information \textit{while} driving where their attention is occupied with limited cognitive capacity available.

\paragraph{Transparency in inequality} Another opportunity in the assistant is to help workers better understand the inequality in existing gig work platforms. With the data collected from the network, the assistant should be able to estimate the level of individual task bias (i.e., whether the pay was equitably decided for an individual worker according to the task allocation---When a worker receives a new task recommendation, the model predicts how much other drivers would be paid for a similar task), group bias (i.e., whether the pay was equitably decided for a demographic group of gig workers according to the task allocation), and the worker-platform bias (i.e., if there exists an alternative task allocation that results in a higher total worker profit with a potentially lower platform profit).\looseness=-1 

%\tlcomment{techniques for measuring AI inequality in gig work}

\subsection{Worker Personal Values}
The core of the end-user programming approach is to allow users to specify their \textit{personal values} in gig work. In any decision making, a \textit{value system} denotes the degree of importance of something for a person, referring to desirable goals and transcending specific actions and situations~\cite{schwartz2012overview}. In a work setting, previous studies have found that different personal values of workers lead to differences in their work values and the priorities in the meaning of work~\cite{ros1999basic} (e.g., good salary, work conditions, authority to make decisions, interesting and varied work, social contact with people). Therefore, in the design of a \textit{personal} intelligent assistant for work planning, it is important that the assistant not only optimizes for a single value (usually the monetary gain), but also considers the weights of other possible personal values (which can vary from one person to another). It is worth noting that a previous study identified that gig workers such as rideshare drivers not only perform body and temporal labor but also emotional labor as a response to the expectation of ``pleasing the passenger'' due to the intermediation of quantitative scores in the rating system~\cite{raval2016standing}.

Two research opportunities are present here. First, more research is needed to empirically understand the work values of gig workers. Many workers value the flexibility in deciding when to work and when to \textit{stop} work with little or no notice. Some rideshare drivers also enjoy the social aspects of their job, e.g., small talk with riders, while others do not~\cite{lutz2018emotional,gloss2016designing}. Additional factors such as e.g., how much does a driver value \textit{not} being stuck in traffic emotionally beyond the value 
of time and gas, how should an intelligent assistant trade-off between a high expected income vs. high expected stability (i.e. low variance) in income may also play a role. Results from such a need-finding study can decide the space of expressiveness that an intelligent assistant should support for gig workers to specify their personal values.\looseness=-1

Another research opportunity is to design interfaces for workers to specify their values. A direct manipulation interface is a possible option, but the user's uncertainty and lack of awareness about their values may pose potential issues. An example-based interface where workers compare tasks of characteristics and choose which ones they prefer can also be useful so the system can infer worker values from their preferences on concrete example tasks. 

% \subsection{Challenges and Opportunities}

% \subsubsection{Measuring and Characterizing AI Inequality in Current Platforms}

\section{Discussion}
\subsection[A ``Fight Fire with Fire'' Approach]{A ``Fight Fire with Fire'' Approach\footnote{We thank the anonymous Reviewer 3 for coming up with this analogy.}}
This position paper advocates for a ``fight fire with fire'' approach---to address the inequality arising from platforms attempting to dictate data access in a centralized top-down fashion and optimizing for platforms' interests in pricing and task allocation decisions using AI models, we seek to democratize data access in a decentralized bottom-up fashion and empower individual works to similarly optimize for their own interests in task planning using AI models.\looseness=-1

Compared with other data-driven efforts that seek to address the same problem, such as Driver's Seat~\cite{drivers_seat_cooperative_drivers_2022}, Shipt~\cite{calacci_bargaining_2022}, and Gigbox~\cite{calacci_gigbox_2022}, our envisioned intelligent agent approach is complementary by providing information and recommendations with a finer granularity and in a more timely fashion. Existing approaches mostly focus on providing \textit{post-hoc} worker support. They often first collect work-related data from workers and later display aggregated statistics (e.g., net earnings, expense per mile, average tips), often in a dashboard, to workers to help them analyze their work records. By showing such data and helping users visualize e.g., how earnings change over time and differ for tasks of different characteristics, these systems can help workers reflect on their work practice and adjust their \textit{higher-level} work strategies to optimize their earnings. Little support is provided to workers to translate such high-level strategies into specific decisions they need to make in the field. In comparison, our envisioned intelligent agent provides real-time recommendations to workers for making \textit{lower-level} specific decisions such as whether to accept the current task and which area to go to when idling. In addition to relying on analytics from aggregated past data like existing ones, our envisioned approach also predicts the \textit{current} market conditions using both historical data and the sample of real-time current data collected from the agent network. \looseness=-1

The complementary relationship between these two types of tools can be found in studies of tooling practices by online gig workers (e.g., crowd workers on platforms such as Amazon Mechanical Turk)~\cite{williams_perpetual_2019}. Workers use both (1) real-time tools that ``automate finding work and accepting work based on specific criteria'' and ``augment the cognition of workers''; and (2) administrative tools that ``track and document work history.''  

The design decision of providing task-level recommendations and utilizing real-time market predictive models exemplifies our ``fight fire with fire'' approach. The platform leverages large-scale real-time data sensing and complicated predictive models to make specific fine-granular decisions on pricing and task allocation. To ``balance the playfield'', we believe that workers ought to be equally equipped with a similar level of technological support.

\subsection{Balancing Worker Agency and Inclusion in Customer Access}
So far we have been mostly discussing the AI inequality between platforms and gig workers with the focus on empowering workers with access to data and AI technologies. However, an important topic, the inclusion in customer access to gig work services, has been left out of the scope of the discussion. When workers have access to AI-enabled assistants that help them with work planning so they are more likely to get the types of task that they \textit{desire}, a possible side effect is that customers whose tasks are considered ``undesirable'' by many workers may be affected by longer wait times, higher prices, or in the worst case, exclusions from the service. 

Note that such biases in service availability for customers have already been identified in existing platform-owned algorithmic systems in previous studies such as~\cite{thebault-spieker_avoiding:_2015,thebault-spieker_toward:_2017}, but the use of algorithmic optimization by workers \textit{might} amplify these biases. Even if the system does not, for example, allow drivers to use protective classes in their configuration (e.g., one \textit{does not} want to serve customers of color), demographic variables can still be proxied using e.g., geographic variables. Therefore, any recommendations and planning assistance provided by the assistant should be validated and balanced for access and inclusion using the predictive task model (i.e., if the use of a specified personal value model would result in a significant shift in the equality of accessing services for customers). Persuasive and de-biasing techniques can also be developed and deployed to reduce unconscious biases and conscious discriminatory practices in gig workers' use of AI assistance.

%\tlcomment{fun things to talk about}

\subsection{Power in Collective Worker Organizations}
% \tlcomment{talk about how this structure may strengthen the collective power of workers through a "AI/data-facilitated organization", and its potential to facilitate collective actions}
In the envisioned approach, a group of workers are connected by data sharing facilitated through each worker's intelligent assistant for a purpose in the collective interest of the group---reducing the AI inequality in gig work. This could be a form of labor organization. The entire group benefits from the participation of each worker, as the contributed data enable more accurate predictive modeling of the market, resulting in better performance for each worker's intelligent assistant in work planning optimization. However, to some extent, gig workers in the same area are also ``competing'' with each other for tasks. It would be interesting to, through community-engaged participatory design processes and deployment studies, investigate worker behaviors in such an organization and how different design decisions made in the system can affect their behaviors.\looseness=-1 

On the optimization aspect, we have so far discussed the assistant's role in helping \textit{individual} workers optimize for their \textit{personal values} in their work planning. However, thanks to the formation of the worker organization, there is also the opportunity for all assistants in the network to coordinate together for the \textit{collective interest} of the group~\cite{vandaele2018will}. Such data/AI-facilitated and ``loosely connected'' (compared with a traditional labor union) groups might still have the potential to increase the bargaining power of workers. For example, if tasks of a specific category (e.g., originated from some regions) are systematically underpaid, the broad use of assistants may significantly cut the labor supply, and subsequently, affect the service availability for these tasks. As a result, the platform, either as a result of automated algorithmic adjustments or manual intervention, may raise the pay for these tasks. The use of intelligent assistant networks can also be used as an infrastructure to facilitate the organization of more formal labor protests and strikes~\cite{flanagan2021can,nissim2021future,maffie2020role}, especially at a finer granularity (e.g., assistants that are configured to recommend declining all tasks that fit specific criteria as part of a labor movement). 

All these interdisciplinary challenges would be a valuable opportunity for the HCI community to engage with sociologists, labor economists, behavioral economists, and researchers from other adjacent disciplines in designing and studying such complex socio-technical systems.

\subsection{Integration with Existing Systems for Immediate Real-World Impacts}
As discussed, the design of the envisioned technical approach prioritizes the integration with existing gig work platforms for immediate real-world impacts. The technology should be compatible with major gig work platforms (e.g., Uber, Lyft) and should incorporate inputs from the community in the design process for real-world use in the actual work context. Specifically, the technical approach assumes only having access to the data from a smaller group of gig workers without requiring any data access provided by the platform operator. \looseness=-1

As a result, there is a lower barrier to making the technology publicly available and planning a field study to (1) evaluate the feasibility, robustness, and ecological validity of the
system features; (2) measure the usefulness of our system in real-life scenarios; and (3) study the characteristics of how users use the system. The real-world adoption of the approach is also more feasible in the near term. In comparison, any approach that requires the cooperation of the platform operator (e.g., developing more equitable pricing and task allocation algorithms for the platform~\cite{chakraborty2017fair,patro2020incremental} and developing an explainable and transparent worker’s app for the platform) is more difficult because the goal of reducing AI inequality in gig work contradicts directly their interests. 

\subsection{Racial Biases in Gig Work}
The rise of the gig economy has particularly affected racially minoritized workers. Empirical research found that ``of survey respondents who earn more than 40\% of their income from on-demand work, a whopping 67\% identify as racial minorities''~\cite{prassl2018humans}.  This is alarming given the fact that when the survey was administrated, racial minorities only made up 36\% of the US population. In other words, in 2018, racial minorities were over-represented in the gig economy by nearly double their actual size in the US population. A source of racial inequality in gig work is racialized surveillance, where workers in racial minority groups, for example, are rated lower, receive more complaints, and earn less tips~\cite{rosenblat2016discriminating,rosenblat2017discriminating,ge2020racial}. Lower ratings and higher customer complaints can affect the tasks assigned to workers and, in some cases, lead to workers being deactivated from the platform. The geographic inequality in rideshare, as identified in~\cite{thebault-spieker_avoiding:_2015,thebault-spieker_toward:_2017}, also contributes to the racial inequality for both workers and customers. 

An emphasis of the envisioned research approach should be investigating how to reduce racial inequality as a part of its effort to bridge the technology divide and the data divide in gig work. In the design process, researchers must ensure the participation of workers from diverse racial, gender, cultural, and SES backgrounds, as well as the representation of their values in design decisions~\cite{dillahunt_2015_promise,harrington_deconstructing_2019}.\looseness=-1

\subsection{Implications for Future Community-Owned Cooperative Platforms}
Another proposed approach to address inequality in gig work from the research community is to design community-owned, worker-owned, or publicly-owned gig work platforms that deviate from the prevailing paradigm where gig work platforms are owned and operated by for-profit corporations (e.g., efforts like The Drivers Cooperative~\cite{the_drivers_cooperative_drivers_2022}, RideFair~\cite{ridefair_ridefair_2022}, and research in platform cooperativism~\cite{scholz2017ours}). In fact, in the space of shared economy in digital products, worker-owned platforms such as Stocksy~\cite{stocksy_united_authentic_2022} (a platform cooperative of photographers) have been shown to be feasible and effective in ensuring equality and fair pay for its members. The open-source software movement and the success of Wikipedia are also examples of success in the decentralized community-owned approach to coordinating, moderating, and managing large-scale group work. However, the adoption of this approach in physical (i.e., offline tasks such as food delivery and rideshare) gig work remains challenging for various reasons, such as the high initial investment needed for setting up the infrastructure, the significant overhead in regulatory compliance, worker background checking, and the challenge in work quality control. Although there have been small-scale successful examples such as Arcade City Austin~\cite{ohalloran_how_2022,arcade_city_austin_arcade_2022} (an online ``forum'' that matches independent drivers with riders through Facebook Page posts), scaling up these efforts seems challenging.\looseness=-1

Our envisioned approach complements the efforts for future community-owned cooperative platforms. Through the deployment of a worker-owned network of intelligent assistants within the ecosystem of corporate-owned gig work platforms, researchers can study worker behaviors, experiment with design strategies, and gain valuable understanding of the current platforms and their inequality in a much shorter timeframe without the barrier of large initial efforts in developing the infrastructure of an entirely new gig work platform and promoting it to reach the critical mass required for study and for the platform to be self-sustainable. The empirical findings, design implications, and interaction strategies learned from our envisioned approach can then be applied to the design and development of future community-owned cooperative platforms.\looseness=-1

\subsection{Broader Impacts in Community Engagement and Labor Advocacy}
This research agenda presents ample opportunities for community engagement and labor advocacy. For example, in the formative stage, we partnered with Chicago Rideshare Advocates~\cite{chicago_gig_alliance,chicago_rideshare_advocates}, a Chicago-based labor advocacy organization with around 3,000 rideshare drivers from a wide range of racial, cultural, and socioeconomic backgrounds. The goal of bridging the AI inequality in gig work aligns with the group's aim of advocating for fairness in ridesharing. Organizations such as Chicago Rideshare Advocates can leverage their deep community roots to play an important role in potential participatory design workshops, data collections, and field deployments. Their access to real gig workers on platforms such as Uber and Lyft would allow researchers to better understand the work contexts on those platforms and ensure the practicality of technological innovations. The data and findings resulting from the research can also support labor advocacy efforts for workers' rights by holding platforms more accountable for their AI algorithms, advocating for policy development, and persuading the companies to make their platforms fairer for the workers.

\section{Conclusion}
This position paper discusses a bottom-up end-user intelligent assistant approach to empower gig workers against AI inequality. We argue that the discussed approach would be a practical way to empower gig workers to bridge the ``data divide'' and ``technology divide'' in gig work that can result in immediate real-world impacts while contributing significant findings and implications for characterizing AI inequality in gig work and designing future cooperative gig work platforms owned by communities. 

The envisioned research direction addresses a complex socio-technical challenge that requires methods, techniques, and theories across multiple disciplines both within computing (e.g., Machine Learning, Data Mining, Human-Computer Interaction) and beyond (e.g., Design, Economics, Psychology, and Sociology). We hope this position paper can spark dialogues spanning multiple disciplines on this increasingly important research topic for the future of work.  

\begin{acks}
This research was supported in part by a Google Cloud Research Credit Grant, a Google Research Scholar Award, and a Lucy Scholar Award. We would like to thank our anonymous reviewers for their feedback. We are grateful to Nitesh Chawla, Daniel Graff, Lori Simmons, Zheng Zhang, Zheng Ning, and Simret Gebreegziabher for useful discussions.
\end{acks}

\balance
\bibliographystyle{ACM-Reference-Format}
\bibliography{references}

\end{document}